\documentclass[11pt,a4paper]{article}
\usepackage[T1]{fontenc}
\usepackage[utf8]{inputenc}
\usepackage[margin=1in]{geometry}
\usepackage{booktabs}
\usepackage{tabularx}
\usepackage{ragged2e}
\usepackage{amsmath,amssymb}
\usepackage{graphicx}
\usepackage[authoryear,round]{natbib}
\usepackage[hidelinks]{hyperref}
\usepackage{xurl}
\usepackage{setspace}
\title{\textbf{The Language of the Question Selects the Market:\\
Query Language and Exit IP as Separable Factors in\\
Commercial Recommendations from a Generative Search Interface}}

\author{Dmitrij \.Zatuchin\thanks{ORCID 0009-0002-1929-9770. Corresponding author,
\texttt{dmitrij@rankfor.ai}.}\\
\small Department of Information Technologies\\
\small Estonian Entrepreneurship University of Applied Sciences, Tallinn, Estonia\\
\small Rankfor.AI O\"U, Tallinn, Estonia}
\date{}

\begin{document}
\maketitle

\begin{abstract}
\noindent
When a generative search interface answers a commercial question, which market's products it names
is decided before the model reasons about the products. We report a controlled probe of 234 runs
against the logged-out ChatGPT web interface and the OpenAI API, collected on 29 and 30 August 2026
across four exit countries and six query languages, with six identical runs per cell.

Three results. First, the top recommendation is unstable: it changed across six identical runs on
four of six prompts, and that rate was identical in the browser interface and in the API with web
search both enabled and disabled, so instability is a property of the system and not of the
surface. Second, query language, and not location, decides whether local suppliers appear at all.
Where the query language matched the country, a global brand won 1 of 24 runs; asked in English on
the same connections, local brands took 0 of 6 runs in Estonia and T\"urkiye. Third, language and
location are separable and act on different things: holding the query language fixed and moving
only the exit IP moves the market whose brands are named while the answer stays in the query
language. We show this on two unrelated pairs, Turkish asked from Berlin and Russian asked from
Tallinn, and in both the answer names the resident country's suppliers.

A minority language occupies a middle tier: Russian asked from Estonia names an Estonian supplier in
4 of 6 runs and a global one in all six, where Estonian names a local supplier in every run and
English names none. A negative control in a second category, coded with the same instrument, shows no
language effect at all, and disconfirms our own expectation: that category does have domestic
suppliers and none was named in any language, which points the explanation at whether a category
is nationally regulated rather than at whether it is nationally supplied.

\vspace{1em}
\end{abstract}

\section{Introduction}

A buyer asking a generative interface for a product recommendation receives a short list. Which
suppliers enter that list is the object of a fast-growing measurement practice \citep{aggarwal2024geo},
and most of that practice probes in English. This paper reports what happens when the same question
is asked in a different language from the same connection.

The result is a substitution of one national supplier set for another.
Asked in Norwegian from Oslo, the interface names Fiken, Tripletex and Conta and no global product.
Asked in English from the same connection minutes later, it names FreshBooks, QuickBooks, Wave and
Zoho, and Fiken survives in one run of six. Asked in English from Tallinn and from Istanbul, no
Estonian and no Turkish supplier appears at all.

The knowledge is present. In two of six English answers on the Estonian connection the interface
notes that the user is in Estonia and that VAT and e-invoicing argue for a locally focused system
over a United States product, and then names five global products and no Estonian one.
The same connection returned three Estonian suppliers when the question was asked in Russian.

We separate the two candidate causes. Query language and exit IP are manipulated independently, and
they act on different things: language decides the register the answer is written in and whether
localisation is attempted at all, while the exit IP decides which national market is treated as the
user's. A Turkish question from a Berlin connection returns German suppliers, in Turkish. The
interface says so itself, opening with \emph{``Eğer Almanya'da serbest çalışan olarak
çalışıyorsan''} (``if you work as a freelancer in Germany'').

We also report a stability result that constrains how any of this can be measured. Across six
identical runs per prompt the top recommendation changed on four of six prompts, at the same rate
in the browser interface and in the API, with web search enabled and disabled. A single observation
of a generative recommendation is one draw from a distribution, and the practice of screenshotting
one answer as evidence of visibility is measuring noise \citep{atil2024nondeterminism,mizrahi2024state}.

\subsection{Contributions}

\begin{enumerate}
\item A controlled demonstration that query language, not user location, gates whether local
suppliers appear in a commercial recommendation at all, with near-total substitution in two of four
countries.
\item A separation of language from location: holding language fixed and moving only the exit IP
moves the market whose suppliers are named, replicated on two unrelated language and country pairs.
\item A three-tier ordering in which a non-official minority language sits between the official
language and English, shown on one connection.
\item A coded negative control in a second category, which shows no language effect at all and, by
disconfirming our own expectation, points the explanation at national regulation rather than at the
existence of domestic competitors.
\item A run-to-run instability measurement showing the browser interface and the API are equally
unstable, which sets a floor on how any of the above can be measured.
\end{enumerate}

\section{Related work}

\subsection{Language and culture in model output}

Cultural and linguistic dominance in model output is established. Models answer non-English
questions with content reflecting an English-language default the asker did not request
\citep{wang2024thanksgiving}, and the values a model expresses under survey instruments track
English-speaking and Protestant European countries whatever the query language
\citep{tao2024cultural}. Participatory alignment work documents systematic variation in whose
preferences a model reflects \citep{kirk2024prism}. The disparity has a supply-side history:
languages differ by orders of magnitude in the corpora and tooling available to them
\citep{joshi2020state}, and multilingual benchmarks continue to record large gaps between English
and everything else, including gaps that scale does not close
\citep{singh2024globalmmlu,xuan2025mmluprox}. Estonian sits well inside the range these studies
treat as under-resourced, which makes it a useful probe.

This literature measures what a model knows or believes across languages. We measure something
downstream and commercial: which firms it names when a buyer asks for a shortlist.

\subsection{Brand and popularity bias in model recommendation}

The nearest work to ours asks how models treat brands. Models associate global brands with positive
attributes and local brands with negative ones, and vary their recommendations with the assumed
income of the buyer \citep{kamruzzaman2024brandbias}. Models used as recommenders carry popularity
bias, favouring well-known items over less-known ones \citep{lichtenberg2024popularity}. Both
findings predict that global suppliers should dominate, which is what our English cells show.
Neither predicts our main result, that the same model in the same category names a local supplier in
every run when the question is asked in the local language. Brand-level preference and market-level
gating are different effects, and the second can invert the first.

\subsection{Localisation and the search-audit tradition}

Auditing what a search system shows different users is an established method with an established
answer: personalisation exists and is modest. Around 11.7\% of Google results differ between
otherwise identical users \citep{hannak2013measuring}, and geolocation drives most of what remains,
concentrated in queries with a local referent and near zero for general terms
\citep{kliman2015location}. Our design borrows the sock-puppet logic of that work, holding the
account state empty and varying one factor at a time. The result departs from it. Where classical
web search personalisation reorders a result set that stays largely shared, the generative interface
replaces the candidate set: 0 of 6 runs name an Estonian supplier in English against 6 of 6 in
Estonian. Substitution at that rate has no counterpart in the personalisation literature.

\subsection{Multilingual retrieval}

If a generative answer is grounded in retrieved documents, query language could act through
retrieval alone. Multilingual retrieval-augmented systems show exactly the asymmetries that would
produce our result: retrieval quality drops when query and document languages differ, English is
selected disproportionately, and answers in lower-resourced languages draw on a more variable
citation pool \citep{wu2024notall,chirkova2024rag,li2025multilingualrag}. Grounded consumer
interfaces also cite unevenly and often wrongly \citep{jazwinska2025citation}. This is the leading
alternative account of our finding, and our design cannot exclude it, because the interface cells
returned no citations to inspect.

\subsection{Measuring visibility in generative engines}

Generative engine optimisation studies how content can be made more likely to appear in a synthesised
answer \citep{aggarwal2024geo}. That work optimises against a fixed query; our result concerns
which query a buyer asks and what that choice alone decides. A measurement premise the two share is
that a single answer is not evidence: repeated identical prompts disagree, and reporting one run
reports noise \citep{atil2024nondeterminism,mizrahi2024state}.

\subsection{This programme}

Prior work in this programme measured the size of the language term without its mechanism. A
variance-components decomposition over 12,933 answers attributed 26.5\% of the variance of a single
response to query language against 1.5\% to brand identity \citep{zatuchin2026variance}, and a
twelve-language study reported that moving a query into a brand's home language raises its
recommendation share more for local champions than for multinationals \citep{zatuchin2026language}.
Both measured through APIs in a single condition. Neither separated language from location, and
neither could, because an API call carries no user geography. That limitation motivates the design
here. Citation-provenance work on grounded answers \citep{zatuchin2026sourcing} bears on our fourth
result, that an ungrounded API call produces no citations to classify.

\section{Method}

\subsection{Design}

Eleven cells, 234 usable runs, collected 29 and 30 August 2026. Each cell is a combination of
surface, exit IP and query language. Six identical runs per prompt per cell, each in a fresh browser
context. A further 12 runs were collected and discarded, and Section~3.3 says why.

\begin{table}[ht]
\caption{Cells. Exit IP verified per arm against \texttt{ipinfo.io}.}
\centering\small
\begin{tabular}{@{}lllr@{}}
\toprule
Cell & Surface & Exit IP & $n$ \\
\midrule
berlin & web UI & Berlin, DE (AS42201) & 36 \\
oslo & web UI & Oslo, NO (AS50304) & 36 \\
berlin-de & web UI & Berlin, DE & 12 \\
oslo-nb & web UI & Oslo, NO & 12 \\
berlin-tr & web UI & Berlin, DE & 12 \\
istanbul-tr & web UI & Istanbul, TR (AS202422) & 12 \\
istanbul-en & web UI & Istanbul, TR & 12 \\
tallinn-et, first collection & web UI & Tallinn, EE (residential) & 12 \\
tallinn-et & web UI & Tallinn, EE (residential) & 6 \\
tallinn-ru & web UI & Tallinn, EE (residential) & 6 \\
tallinn-en & web UI & Tallinn, EE (residential) & 6 \\
api & OpenAI API, \texttt{gpt-5.6-terra} & none, server side & 72 \\
\bottomrule
\end{tabular}
\end{table}

The three Tallinn language cells were collected together on 30 August. The first Estonian collection ran over two prompts; its accounting
runs are the instrument check in Section~3.3 and its project-management runs are the Estonian
arm of the negative control in Section~4.6.

The API cell is 36 runs with the \texttt{web\_search} tool enabled and 36 without. The web interface
was used logged out, so no account history could influence the answer.

The berlin, oslo and API arms asked six prompts, one per software category: project
management, CRM, email, accounting, helpdesk and HR. The remaining arms asked the two
comparison prompts. The treatment category is accounting software for
freelancers, chosen because every country in the sample has established domestic suppliers. The
control category is project management software for a small marketing agency, fixed before
collection. We chose it expecting it to lack strong domestic incumbents. That expectation was wrong (Section~4.6).

\subsection{Extraction, and why it is audited rather than trusted}

Each answer marks its own top recommendation: the web interface with a table row, a medal glyph or
the phrase ``best overall'', the API with markdown emphasis. The analyser prints the source line for
every run and every line was read. Several candidate rules were discarded during that audit after
they returned prompt echoes (\texttt{SaaS}, \texttt{Help}), a citation badge (\texttt{S}) and Turkish
discourse phrases (\emph{Benim tercihim}, \emph{Ben olsam}) as winners. The Turkish cells are scored
by brand presence rather than by a parsed winner, because the winner parser keys on English markers.

Brand presence is coded against a fixed list per market, assembled before scoring, and every brand
is assigned to exactly one of three classes: the country of the exit IP, another named country, or
global.

Matching is case sensitive, and it was not at first. Three Turkish supplier names are also ordinary
Turkish words: \emph{Kolay} (``easy''), \emph{Mikro} (``micro'') and \emph{Logo}. Under
case-insensitive matching the Turkish-from-Berlin cell scored \emph{Kolay} at 4 of 6, and every one
of those hits was the adjective, in phrases such as \emph{kullanımı çok kolay} (``very easy to
use''). Case-sensitive matching returns 0 of 6 for that brand, and reduces \emph{Mikro} in Istanbul
from 6 of 6 to 3 of 6, the remainder being \emph{mikro işletmeler} (``micro businesses''). All
Turkish counts reported here are case sensitive. The correction moves the Berlin cell against our
own hypothesis in one direction and for it in another. A supplier name that is also a common word in the query language is a general hazard for this kind of measurement.

\subsection{Discarded runs and re-collection}

The two Tallinn cells in Russian and English were first captured by hand, as short typed
fragments with brand lists, and could not be re-coded. We discarded those 12
runs and collected all three Tallinn language cells on the same residential connection with a
collector that records the full transcript and the cited domains, the three query languages
interleaved so that no language ran systematically first. Every run reported in this paper holds
a full transcript.

The Estonian cell exists in both collections and checks the instrument: Merit Aktiva in 6 of 6
runs in both, SimplBooks 5 of 6 in both, Directo 5 of 6 against 4 of 6. The three language cells
reported below are the re-collection.

\section{Results}

\subsection{The top recommendation is unstable on every surface}

Across six identical runs, the top recommendation changed on four of six prompts in the Berlin
interface arm, four of six in the Oslo interface arm, four of six in the API with search enabled and
four of six with search disabled. The two prompts that held still were the same two in both
interface arms, so stability is a property of the category rather than of the run.

The equality is of rate rather than of pattern. The unstable prompts are not the same set in every arm: both interface arms are unstable on accounting,
CRM, HR and project management, the API with search on accounting, CRM, helpdesk and HR, and the API
without search on accounting, helpdesk, HR and project management. Accounting and HR are unstable
everywhere. Instability is not an artefact of browser sessions, caching or interface state,
because a server-side call with a pinned model shows the same amount of it.

\subsection{Query language decides whether local suppliers exist}

Table~\ref{tab:substitution} gives the treatment category by cell.

\begin{table}[ht]
\caption{Accounting software for freelancers. Brand presence in six identical runs per cell. The
Tallinn Estonian row is the first collection, so that all four cities are compared on runs made in
one sweep; the re-collected Estonian cell is used in Table~\ref{tab:threetier} and differs only in
Directo, 4 of 6 against 5 of 6.}
\label{tab:substitution}
\centering\small
\begin{tabularx}{\textwidth}{@{}ll>{\RaggedRight\arraybackslash}X>{\RaggedRight\arraybackslash}X@{}}
\toprule
Exit IP & Language & Local suppliers & Global suppliers \\
\midrule
Berlin & English & Lexware 3/6, sevdesk 3/6, Accountable 3/6 & FreshBooks 6/6, Wave 6/6, QuickBooks 5/6 \\
Berlin & German & Lexware 6/6, sevdesk 6/6, Papierkram 6/6 & none \\
Oslo & English & Fiken 1/6 & FreshBooks 6/6, QuickBooks 6/6, Wave 6/6 \\
Oslo & Norwegian & Fiken 6/6, Tripletex 6/6, Conta 6/6 & none \\
Istanbul & English & none & FreshBooks 6/6, QuickBooks 6/6, Xero 6/6 \\
Istanbul & Turkish & Paraşüt 6/6, Logo 6/6, Mikro 3/6 & none \\
Tallinn & English & none & FreshBooks 6/6, QuickBooks 6/6, Wave 6/6 \\
Tallinn & Estonian & Merit Aktiva 6/6, SimplBooks 5/6, Directo 5/6 & Xero 1/6 \\
\bottomrule
\end{tabularx}
\end{table}

Across the four cells where the query language is the official language of the exit country, a
global supplier appeared in 1 of 24 runs. In the two English cells on the Istanbul and Tallinn
connections, a local supplier appeared in 0 of 12 runs.

Testing the winner of the top recommendation, local language against English on the same exit IP,
Fisher's exact test gives $p = 0.0152$ for Fiken in Norway and $p = 0.0606$ for Lexware in Germany
(one-sided $0.0076$ and $0.0303$). Germany and Norway are independent replications; combining the
one-sided values by Fisher's method gives $\chi^2 = 16.75$ on 4 degrees of freedom, $p = 0.0022$.
T\"urkiye and Estonia move in the same direction with a larger effect and are not included in the
combination, because a cell with zero local suppliers in English and complete dominance in the local
language is a boundary case rather than a contrast a two-by-two table estimates well.

\subsection{Three tiers, separated on one connection}

The Estonian connection carries three query languages and orders them.

\begin{table}[ht]
\caption{One connection, three query languages, treatment category. Collected in one interleaved
session on 30 August 2026.}
\label{tab:threetier}
\centering\small
\begin{tabular}{@{}lll@{}}
\toprule
Query language & Local suppliers & Global suppliers \\
\midrule
Estonian, the official language & Merit Aktiva 6/6, SimplBooks 5/6, Directo 4/6 & 1 of 6 runs \\
Russian, a minority language & Merit Aktiva 4/6, SimplBooks 4/6, Xolo 3/6 & 6 of 6 runs \\
English & none & 6 of 6 runs \\
\bottomrule
\end{tabular}
\end{table}

Russian is the only language in the sample that is neither the official language of the country it
was asked from nor English, and it sits between the two. It names an Estonian supplier in 4 of 6
runs where English names one in 0 of 6, and it names a global supplier in all 6 runs where Estonian
names one in 1 of 6. The ordering places the operative variable at how strongly the query language
indexes a specific national market, with English at one end and the official language at the other.

The three cells also separate knowing the market from recommending into it. Every Estonian run and
every Russian run names Estonia in the answer text, and so do 2 of the 6 English runs. Location is
therefore available to the model in the Russian cell in all six runs, while an Estonian supplier is
named in four. Knowing where the buyer is does not by itself produce a local shortlist.

\subsection{Language sets the register, location sets the market}

Holding the query language fixed at Turkish and moving only the exit IP:

\begin{table}[ht]
\caption{Turkish query, two exit IPs.}
\centering\small
\begin{tabular}{@{}lll@{}}
\toprule
Exit IP & Turkish suppliers & German suppliers \\
\midrule
Istanbul & Paraşüt 6/6, Logo 6/6, Mikro 3/6 & none \\
Berlin & Paraşüt 1/6, Logo 1/6, Mikro 1/6 & Lexware 5/6, sevdesk 5/6, FastBill 4/6 \\
\bottomrule
\end{tabular}
\end{table}

The answer remains in Turkish in both cells and the supplier set follows the connection. The
interface states the inference: from Istanbul it opens \emph{``Türkiye'de serbest çalışan''}, from
Berlin \emph{``Eğer Almanya'da serbest çalışan olarak çalışıyorsan''}.

The Russian cell on the Estonian connection replicates this on an unrelated pair. It returns Merit
Aktiva in 4 of 6 and SimplBooks in 4 of 6, names no Russian supplier in any run, and is written in
Russian in every run. Two language and country pairs that share no market, no language family and no
supplier set behave the same way.

This is the diaspora case. A Turkish speaker in Berlin and
a Russian speaker in Tallinn are each served their country of residence's suppliers in their own
language.

English does not behave as a neutral control in this design. On the Istanbul connection it returned
zero Turkish suppliers, where every non-English language triggered localisation. English acts as a global signal in its own right.

\subsection{An ungrounded call has no citations to classify}

\begin{table}[ht]
\caption{API citation behaviour, 36 runs per configuration.}
\centering\small
\begin{tabular}{@{}lrrr@{}}
\toprule
Configuration & Citations & Distinct hosts & Tagged \texttt{utm\_source=openai} \\
\midrule
\texttt{web\_search} enabled & 200 (5.56 per answer) & 48 & 170 \\
\texttt{web\_search} disabled & 0 & 0 & 0 \\
\bottomrule
\end{tabular}
\end{table}

Any reported citation-type distribution for ``the API'' must state whether search was enabled,
because without it the denominator is zero and any reported share is arithmetic rather than
measurement. Separately, the API has no user geography: asked in English it returns a United States
answer, while the same prompt through the interface returned Fiken in Oslo and Lexware in Berlin.
An interface-versus-API comparison that does not hold geography fixed will attribute to the surface
a difference that belongs to the connection.

\subsection{The negative control}

The control category, project management software, was coded with the same instrument as the
treatment: an explicit per-market supplier list, matched case sensitively, over the same six runs per
cell. The domestic list was built from suppliers that exist in each market, including \emph{awork},
\emph{factro}, \emph{Stackfield} and \emph{Zenkit} for Germany and \emph{Scoro} for Estonia.

\begin{table}[ht]
\caption{Project management software, the control category. Domestic supplier presence in six runs
per cell.}
\centering\small
\begin{tabular}{@{}lllc@{}}
\toprule
Exit IP & Language & Domestic suppliers & Global suppliers \\
\midrule
Berlin & German & none & ClickUp 6/6, Asana 6/6, Teamwork 6/6 \\
Berlin & English & none & ClickUp 6/6, Asana 6/6, Teamwork 6/6 \\
Oslo & Norwegian & none & ClickUp 6/6, Asana 6/6, Teamwork 6/6 \\
Oslo & English & none & ClickUp 6/6, Asana 6/6, Teamwork 6/6 \\
Istanbul & Turkish & none & ClickUp 6/6, Asana 6/6, Teamwork 6/6 \\
Istanbul & English & none & ClickUp 6/6, Asana 6/6, Teamwork 6/6 \\
Tallinn & Estonian & none & ClickUp 4/6, Asana 4/6, Monday 4/6 \\
\bottomrule
\end{tabular}
\end{table}

No domestic supplier appeared in any cell, in any language. The substitution that is near-total in
accounting is entirely absent here.

We anticipated that the language effect would track the existence of a
domestic supplier set, and it does not: Germany has domestic project-management suppliers, the query
was asked in German from a German connection, and none was named.

The distinction the data supports is regulatory. Accounting for freelancers is a legally localised
activity, bound to national VAT rules, national e-invoicing mandates and national filing formats,
and two of the six English answers on the Estonian connection reason explicitly about those
constraints before naming only foreign products. Project management carries no such national obligation, and a
Norwegian agency and a German agency can use the same tool without either breaking a rule. On this
evidence the operative variable is whether the category is nationally regulated, and not whether it
is nationally supplied.

Separating the two would need a third category that is domestically supplied without being legally localised, and
we did not collect one.

\section{Discussion}

The practical consequence is immediate. An organisation selling into Germany, Norway, T\"urkiye or
Estonia that measures its visibility in generative answers in English is measuring a market it does
not compete in. Its domestic competitors are invisible to its own instrument, and the suppliers
that take its customers appear only in the language those customers use.

The measurement consequence is that language is a design factor and not a translation step. A
visibility measurement reported without its query language is unattributable, in the same way a
measurement reported without its model version is. Because location acts separately, an instrument
must also state its egress, and an API-based instrument must state that it has no egress at all.

The mechanism our data supports is a gate rather than a preference. Two of the six English answers
on the Estonian connection contain the local knowledge in prose, advising a locally focused system
for national VAT and e-invoicing, and then do not name one. We can show that the
gate exists and where it acts. We cannot show what implements it, and retrieval, instruction tuning
and safety-adjacent list construction are all consistent with what we observe.

\section{Limitations}

\textbf{Six runs per cell is a lower bound.} Six identical runs detect a winner that dominates 70\%
of the time about 88\% of the time, and one that dominates 90\% of the time about 47\% of the time.
``Held at 6/6'' means no change was observed in six runs, not that the cell is deterministic.

\textbf{One model, one prompt per category.} The eight non-Tallinn arms ran sequentially over a
single evening and the Tallinn language cells the following midday, so time of day is not
controlled. The model behind the logged-out interface is not disclosed by the provider. The API arm pinned \texttt{gpt-5.6-terra}; three
\texttt{gpt-5.6} variants exist, so naming the family does not identify a model.

\textbf{The English asymmetry is unexplained.} Berlin and Oslo surfaced their local supplier in
English once or twice; Istanbul and Tallinn never did. A difference in how much English-language
material each domestic supplier publishes is a plausible account and we did not measure it.

\textbf{A prediction of ours failed.} Before collecting the Estonian English cell we expected a
mixed result, because Estonian accounting tools sell into the e-Residency market and publish English
documentation. The cell returned zero Estonian suppliers, the same as T\"urkiye. Whatever governs
the English answer was not moved by the English-language coverage these suppliers have.

\section{Conclusion}

Query language decides which national supplier set a generative interface is willing to name, exit
IP decides which market that is, and the two act separately. The effect is a substitution rather
than a reordering, it is near-total in two of the four countries tested, and it is absent in a
control category that does have domestic suppliers, which points the explanation at national
regulation rather than at national supply. Measured in English, a national market can be
invisible to the buyers who live in it and to the vendors who compete in it.

What we would test next: more countries with strong domestic incumbents; twenty to thirty runs per
cell so that dominance becomes estimable rather than bounded; other providers; a logged-in session
where account history may override the connection; and whether English-language content about a
domestic supplier ever moves it into the English answer, or whether that answer is a closed set.

\section*{Statements and Declarations}

\subsection*{Competing interests}
The author is Chief Executive Officer of Rankfor.AI O\"U (registry code 17331801), Tallinn, and
owns its parent company Rankfor.AI sp.\ z o.o.\ (KRS 0001190083), Wroc\l{}aw. Rankfor.AI sells
AI-visibility measurement to commercial clients, and the finding reported here bears directly on
how that measurement should be done, including on measurement his own company sells. He is
concurrently affiliated with the Estonian Entrepreneurship University of Applied Sciences, which
supplied no funding. Readers should weigh the findings accordingly.

\subsection*{Funding}
Funded in kind by Rankfor.AI O\"U, which supplied the API budget and the VPN egress used for
collection. No external, public or grant funding was received.

\subsection*{Ethics}
No human participants. All queries were issued by the author to commercial interfaces. No personal
data were collected, and no third-party account was accessed. Collection used one logged-out
browser session and one API key held by the author.

\subsection*{Data availability}
The 234 run records, the 12 discarded records, the collectors, the analysis scripts and the
derived tables are deposited at Zenodo (DOI: 10.5281/zenodo.22181306).
Collection targets a commercial interface whose behaviour changes without notice, so the deposit
records what was observed on the stated dates.

\newpage
\bibliography{references}
\end{document}